# A Compact Planar Triple-Nuclear Coil for Small Animal $^1$H, $^{13}$C, and $^{31}$P Metabolic MR Imaging at 14.1 T

A. P. Leynes[1*], Y. Chen[1], S. Sukumar[1], D. Xu[1,2], X. Zhang[1,2]

[1] Department of Radiology and Biomedical Imaging, University of California San Francisco, San Francisco, CA, USA

[2] UC Berkeley and UCSF Joint Graduate Group in Bioengineering, San Francisco and Berkeley, CA, USA

\* Corresponding author

Andrew P. Leynes

Department of Radiology and Biomedical Imaging

University of California San Francisco

Byers Hall 102D

1700 4th Street

San Francisco, CA 94158

Andrew.Leynes@ucsf.edu

**Abstract**

Multi-nuclear radio-frequency (RF) coils at ultrahigh field strengths are challenging to develop due to the high operating frequency, increased electromagnetic interaction among the coil elements, and increased electromagnetic interaction between the coils and the subject. In this work, a triple-nuclear surface coil for $^1$H, $^{13}$C, and $^{31}$P for small animal metabolic imaging at 14.1 T is built by nesting three surface coil geometries: a lumped-element L-C loop coil, a butterfly coil, and a microstrip transmission line (MTL) resonator. The loop coil, butterfly coil, and MTL are tuned to $^{31}$P, $^{13}$C, and $^1$H, respectively. The successful bench tests and phantom imaging with this novel triple-nuclear coil demonstrates the feasibility of the design for triple-nuclear MR imaging and spectroscopy studies at the ultrahigh field of 14.1 T. In particular, the microstrip transmission line resonator demonstrates excellent coil performance for proton imaging at 600 MHz, where conventional lumped-element design is limited.

**Introduction**

Magnetic resonance imaging (MRI) at ultrahigh field strengths increases signal-to-noise (SNR) and allows for greater spatial and spectral resolution. In particular, hetero-nuclei or X-nuclei imaging becomes more viable due to the increases in the Larmor frequency (thus spectral dispersion) and SNR. However, radiation losses, dielectric losses, and electromagnetic coupling effects increase with higher operating frequencies, which leads to technical challenges in designing efficient RF hardware to realize the advantages of ultrahigh field MRI [1]–[31].

In hetero-nuclear MR, multi-nuclear RF coils for MR imaging excitation and reception are often needed, offering capabilities of imaging multiple nuclei in a single experiment without having to reposition the subject. This has the advantage of obtaining structural and metabolic information in living systems and offers precise co-registration between the different imaged nuclei. Numerous double-nuclear coils of different configurations and different designs have been proposed and investigated ($^1$H/$^{13}$C [13], [18], [19], [31], [32], $^1$H/$^{23}$Na [19], [26], [30], [33], [34], and $^1$H/$^{31}$P [27]–[29], [35]–[37]) as well as a few triple-nuclear coils [38], [39]. In the ultrahigh field MR imaging applications, the design challenges of developing multi-nuclear RF coils mainly result from the high operating frequencies and increased electromagnetic coupling between different nuclear channels. Insufficient isolation between nuclear channels can lead to diminished excitation and acquisition performance, particularly in hetero-nuclear MR experiments [40]–[43].

In this work, we conceptualize, develop, and test a simple and compact design for a triple-nuclear coil for the metabolic imaging applications at the ultrahigh field of 14.1T. Standard RF bench tests and phantom MR imaging experiments have been performed to validate the proposed triple-nuclear coil design for each of the selected nuclei.

**Methods**

*Coil Design*

The triple-nuclear surface coil is built using by nesting three surface coil geometries: a lumped-element LC loop coil, a butterfly coil [8], [40], [41], [44]–[48], and a half-wavelength (λ/2) microstrip transmission line (MTL) [1], [2], [4], [4], [6], [9], [10], [12], [15], [18], [20], [21], [25], [31], [49]–[52]. The three coil elements are arranged as shown in Figure 1.

Based on the coil layout, both the MTL and butterfly coil have orthogonal magnetic fields with the loop coil, generating zero net flux and provides excellent decoupling performance [31], [41]. However, the MTL and butterfly coil have parallel magnetic fields and are coupled. Since tuned resonators are inherently band-pass filters [53], tuning the MTL and butterfly coil with a large frequency separation provides the necessary electronic decoupling needed for sufficient operation.

*Coil Construction*

The MTL strip conductor (7 mm × 45 mm × 36 µm) and the ground plate (50 mm × 45 mm × 36 µm) is constructed with adhesive-backed copper tape on a PTFE substrate (45 mm × 45 mm × 7 mm). Additional length (65 mm) of the PTFE substrate was used to extend the ground plane and provide mechanical support for the coaxial cables. Round copper wire (18 AWG, 1.0 mm diameter) was used to construct the loop coil and butterfly coil. The MTL, loop coil, and butterfly coil, were tuned for $^1$H (600MHz), $^{31}$P (243MHz), and $^{13}$C (150MHz), respectively. Non-magnetic fixed and variable trimmer capacitors (Johanson, Camarillo, CA, USA) were used for tuning and matching. The coaxial cables were recessed into the PTFE substrate by 2 mm and mechanically secured with plastic cable ties. The constructed coil is shown in Figure 2.

*Bench Testing*

Bench-tests, tuning, and matching were done using a network analyzer (Agilent Technologies, Model E5061A). Scattering parameters, $S_{11}$ and $S_{21}$, and coil efficiency (Q-factor) were measured for both unloaded and loaded cases for each of the coil elements.

*Phantom Imaging*

Two phantoms were imaged using a Varian 14.1 T vertical bore micro-imaging scanner in separate scan sessions: a 2.5cm diameter sphere containing 5 M enriched $^{13}$C-urea, and a 1.5 cm diameter cylindrical sample tube containing 500 mM sodium phosphate with a pH of 9.2. Separate phantom experiments were performed in order to avoid significant shimming issues when scanning with two geometrically distinct phantoms in a single experiment.

The $^{13}$C-urea phantom was imaged using a 2D gradient-echo sequence (TE = 1.92 ms, TR = 400 ms, α = 40°, $N_{averages}$ = 4, slice thickness = 4 mm, FOV = 40 mm × 40 mm, matrix size = 128 × 128) for the proton image, and a 2D chemical shift imaging (CSI) sequence (TE = 0.53 ms, TR = 500 ms, $N_{averages}$ = 2, spectral points = 256, spectral width = 4kHz, resolution = 2.5 mm × 2.5 mm, slice thickness = 5 mm, FOV = 40 mm × 40 mm) to obtain the $^{13}$C spectra.

The phosphate phantom was imaged using a 2D gradient-echo sequence (TE = 1.88 ms, TR = 30 ms, α = 20°, $N_{averages}$ = 1, slice thickness = 2 mm, matrix size = 128 × 128, FOV = 50 mm × 50 mm) to obtain the proton image, and a non-selective hard pulse (pulse width = 180 $\mu$s, α = 90°, TR = 6s, $N_{averages}$ = 16, complex points = 10000, block size = 8, spectral width = 10kHz, acquisition time = 984ms) to obtain the $^{31}$P spectrum.

*Image Analysis*

Images and spectral data are reconstructed using SIVIC [54]. The $^{13}$C and $^{31}$P spectral data were apodized using exponential filters to give line-broadening factors of 9 Hz and 20 Hz, respectively. Image analysis is performed using MATLAB (The MathWorks, Inc., Natick, MA, USA) and line profiles are measured with the Imagine toolkit [55].

The peak SNR (PSNR) is measured for each image and line spectra. The PSNR is measured by taking the maximum signal and dividing by the standard deviation of the background signals. For MRSI, the spectral SNR is measured for each voxel by taking the integral under the peak; the PSNR is taken from the voxel with the highest SNR. Triangle thresholding [56] is used to differentiate real signal vs. background noise.

## Results

*Bench Test Results*

S-parameter measurements are summarized in Table 1. The coil achieves a worst-case $S_{11}$ of -31.14 dB and $S_{21}$ of -16.41 dB indicating good matching and decoupling performance. Plots of the $S_{11}$ and $S_{21}$ measurements are shown in Supplementary Figure S1. Q factor measurements and tuning ranges are listed in Tables 2 and 3.

**Table 1**. S-parameter matrix

| Coil channel | $^1$H | $^{13}$C | $^{31}$P |
|---|---|---|---|
| $^1$H | -47.94 dB | -16.41 dB | -20.20 dB |
| $^{13}$C | | -42.03 dB | -29.98 dB |
| $^{31}$P | | | -31.14 dB |

**Table 2**. Q factor measurements for each coil channel. The loaded case was simulated by placing the coil on the thigh of a human volunteer.

| Coil channel | $Q_{unloaded}$ | $Q_{loaded}$ | $Q_{ratio}$ |
|---|---|---|---|
| $^1$H | 142 | 48 | 2.96 |
| $^{13}$C | 132 | 28 | 4.71 |
| $^{31}$P | 136 | 40 | 3.40 |

**Table 3**. Tuning ranges of each coil channel.

| Coil channel | Minimum tuned frequency (MHz) | Maximum tuned frequency (MHz) | Tuning range (MHz) | Capacitor tuning range (pF) |
|---|---|---|---|---|
| $^1$H | 560 | 945 | 385 | 2.5 – 10 |
| $^{13}$C | 138 | 226 | 88 | 1.0 – 5.0 |
| $^{31}$P | 197 | 307 | 110 | 1.0 – 5.0 |

*Phosphorus Phantom Experiment*

Results of the phosphorus phantom experiment are shown in Figure 3. The line profiles display extensive sample coverage, as expected from a MTL resonator. The peak SNR for the phosphorus proton image is 222.98 or 23.48 dB. The spectral SNR is 8979 or 39.53 dB for the raw spectrum, and 16062 or 42.06 dB for the apodized spectrum. The experiment demonstrates the feasibility of proton imaging and phosphorus spectroscopy despite the presence of the additional $^{13}$C coil.

*C-13 Phantom Experiment*

Results of the carbon-13 phantom experiment are shown in Figure 4. The peak SNR for the carbon-13 proton image is 451.85 or 26.55 dB. The spectral peak SNR is 373.32 or 25.72 dB. As with the phosphorus phantom experiment, $^{13}$C spectroscopy and imaging were successfully performed with the proposed coil design.

**Discussion**

This study explores a triple-nuclear surface coil design for $^1$H/$^{13}$C/$^{31}$P small animal imaging at the ultrahigh field of 14.1T and demonstrates the feasibility of the proposed coil design. Development of a triple-nuclear coil for 14.1T applications is challenging because of: (1) effects of ultrahigh magnetic field strengths, and (2) coupling between multiple channels. Proton imaging with conventional lumped-element design is limited by the high operating frequency (600 MHz) at 14.1 T, whereas with the MTL resonator design, we achieve high SNR, high Q-factor, and extensive sample coverage. Moreover, the configuration of the different coil geometries provides sufficient decoupling with basic tuning and matching circuitry. Our simple design alleviates the stated technical challenges and provides satisfactory acquisition performance for the specified application.

The coil coverage of C-13 channel is limited to a depth of 5 - 7 mm whereas the proton channel achieves coverage over the whole sample. At low operating frequencies, the coil sensitivity profile is concentrated near the coil; the coil sensitivity profiles become more structured and penetrate deeper into the sample as the operating frequency is increased [7]. Careful design of the butterfly coil should be done in order to achieve sufficient coil sensitivity at a desired imaging depth.

The triple-nuclear coil allows for simultaneous experiments of $^{13}$C and $^{31}$P, which opens up avenues of research into the correlation of different metabolites with these nuclei such as for pH imaging [57]–[59] or for cell energetics [37], [60]. The coil may also be designed to work with other combinations of nuclei, such as $^{1}$H/$^{23}$Na/$^{39}$K, provided that the butterfly coil tuning has sufficient frequency separation with the MTL resonator tuning. Given the simple and compact structure, this triple-nuclear coil design can be a potential design block for multi-channel triple-nuclear RF coil array design if the appropriate broadband decoupling mechanism, such as ICE decoupling technique [61], is applied.

**Conclusion**

A design for a triple-nuclear coil for $^1$H, $^{13}$C, and $^{31}$P small animal imaging at 14.1 T was successfully constructed and tested with the standard RF testing methods and MR imaging experiments. High Q-factors, extensive proton coil coverage, and sufficient decoupling for simultaneous $^1$H, $^{13}$C, and $^{31}$P imaging can be achieved with the proposed design. In particular, the microstrip transmission line resonator demonstrates excellent coil performance for proton imaging at the ultra-high frequency of 600 MHz, where conventional lumped-element design is limited. Improving the basic design of the triple-nuclear surface coil is straightforward: adding dedicated filtering circuits to increase decoupling between the coil channels. Moreover, assembly of a triple-nuclear phased array is feasible due to the compact planar design of the surface coil, provided that the appropriate decoupling method is applied between blocks. The coil design may also be adapted for lower field strengths, such as 9.4 T or 7 T.

**Acknowledgements**

This work was supported in part by National Institutes of Health (NIH) grants R01EB008699, R21EB020283, R01EB012031, and P41EB013598, and a UCSF Academic Senate Award to XZ.

## References

[1] X. Zhang, K. Ugurbil, and W. Chen, "Microstrip RF surface coil design for extremely high-field MRI and spectroscopy," *Magn. Reson. Med.*, vol. 46, no. 3, pp. 443–450, Sep. 2001.

[2] X. Zhang, K. Ugurbil, and W. Chen, "A microstrip transmission line volume coil for human head MR imaging at 4 T," *J. Magn. Reson.*, vol. 161, no. 2, pp. 242–251, Apr. 2003.

[3] J. t. Vaughan, G. Adriany, C. j. Snyder, J. Tian, T. Thiel, L. Bolinger, H. Liu, L. DelaBarre, and K. Ugurbil, "Efficient high-frequency body coil for high-field MRI," *Magn. Reson. Med.*, vol. 52, no. 4, pp. 851–859, Oct. 2004.

[4] G. Adriany, P.-F. Van de Moortele, F. Wiesinger, S. Moeller, J. P. Strupp, P. Andersen, C. Snyder, X. Zhang, W. Chen, K. P. Pruessmann, P. Boesiger, T. Vaughan, and K. Uğurbil, "Transmit and receive transmission line arrays for 7 Tesla parallel imaging," *Magn. Reson. Med.*, vol. 53, no. 2, pp. 434–445, Feb. 2005.

[5] G. C. Wiggins, A. Potthast, C. Triantafyllou, C. J. Wiggins, and L. L. Wald, "Eight-channel phased array coil and detunable TEM volume coil for 7 T brain imaging," *Magn. Reson. Med.*, vol. 54, no. 1, pp. 235–240, Jul. 2005.

[6] X. Zhang, X.-H. Zhu, and W. Chen, "Higher-order harmonic transmission-line RF coil design for MR applications," *Magn. Reson. Med.*, vol. 53, no. 5, pp. 1234–1239, May 2005.

[7] F. Wiesinger, P.-F. Van de Moortele, G. Adriany, N. De Zanche, K. Ugurbil, and K. P. Pruessmann, "Potential and feasibility of parallel MRI at high field," *NMR Biomed.*, vol. 19, no. 3, pp. 368–378, May 2006.

[8] F. D. Doty, J. Kulkarni, C. Turner, G. Entzminger, and A. Bielecki, "Using a cross-coil to reduce RF heating by an order of magnitude in triple-resonance multinuclear MAS at high fields," *J. Magn. Reson.*, vol. 182, no. 2, pp. 239–253, Oct. 2006.

[9] W. Driesel, T. Mildner, and H. E. Möller, "A microstrip helmet coil for human brain imaging at high magnetic fields," *Concepts Magn. Reson. Part B Magn. Reson. Eng.*, vol. 33B, no. 2, pp. 94–108, Apr. 2008.

[10] B. Wu, C. Wang, D. A. C. Kelley, D. Xu, D. B. Vigneron, S. J. Nelson, and X. Zhang, "Shielded Microstrip Array for 7T Human MR Imaging," *IEEE Trans. Med. Imaging*, vol. 29, no. 1, pp. 179–184, Jan. 2010.

[11] C. Constantinides, S. Angeli, S. Gkagkarellis, and G. Cofer, "Intercomparison of performance of RF coil geometries for high field mouse cardiac MRI," *Concepts Magn. Reson. Part A*, vol. 38A, no. 5, pp. 236–252, Sep. 2011.

[12] Y. Pang, Z. Xie, Y. Li, D. Xu, D. Vigneron, and X. Zhang, "Resonant Mode Reduction in Radiofrequency Volume Coils for Ultrahigh Field Magnetic Resonance Imaging," *Materials*, vol. 4, no. 8, pp. 1333–1344, Jul. 2011.

[13] Y. Pang, X. Zhang, Z. Xie, C. Wang, and D. B. Vigneron, "Common-Mode Differential-Mode (CMDM) Method for Double-Nuclear MR Signal Excitation and Reception at Ultrahigh Fields," *IEEE Trans. Med. Imaging*, vol. 30, no. 11, pp. 1965–1973, Nov. 2011.

[14] M. Vossen, W. Teeuwisse, M. Reijnierse, C. M. Collins, N. B. Smith, and A. G. Webb, "A radiofrequency coil configuration for imaging the human vertebral column at 7 T," *J. Magn. Reson.*, vol. 208, no. 2, pp. 291–297, Feb. 2011.

[15] P. Yazdanbakhsh and K. Solbach, "Microstrip Butler matrix design and realization for 7 T MRI," *Magn. Reson. Med.*, vol. 66, no. 1, pp. 270–280, Jul. 2011.

[16] B. van den Bergen, D. W. J. Klomp, A. J. E. Raaijmakers, C. A. de Castro, V. O. Boer, H. Kroeze, P. R. Luijten, J. J. W. Lagendijk, and C. A. T. van den Berg, "Uniform prostate imaging and spectroscopy at 7 T: comparison between a microstrip array and an endorectal coil," *NMR Biomed.*, vol. 24, no. 4, pp. 358–365, May 2011.

[17] K. M. Gilbert, J.-G. Belliveau, A. T. Curtis, J. S. Gati, L. M. Klassen, and R. S. Menon, "A conformal transceive array for 7 T neuroimaging," *Magn. Reson. Med.*, vol. 67, no. 5, pp. 1487–1496, May 2012.

[18] Y. Pang, Z. Xie, D. Xu, D. A. Kelley, S. J. Nelson, D. B. Vigneron, and X. Zhang, "A dual-tuned quadrature volume coil with mixed λ/2 and λ/4 microstrip resonators for multinuclear MRSI at 7 T," *Magn. Reson. Imaging*, vol. 30, no. 2, pp. 290–298, Feb. 2012.

[19] C. Wang, Y. Li, B. Wu, D. Xu, S. J. Nelson, D. B. Vigneron, and X. Zhang, "A practical multinuclear transceiver volume coil for in vivo MRI/MRS at 7 T," *Magn. Reson. Imaging*, vol. 30, no. 1, pp. 78–84, Jan. 2012.

[20] B. Wu, C. Wang, J. Lu, Y. Pang, S. J. Nelson, D. B. Vigneron, and X. Zhang, "Multi-Channel Microstrip Transceiver Arrays Using Harmonics for High Field MR Imaging in Humans," *IEEE Trans. Med. Imaging*, vol. 31, no. 2, pp. 183–191, Feb. 2012.

[21] B. Wu, X. Zhang, C. Wang, Y. Li, Y. Pang, J. Lu, D. Xu, S. Majumdar, S. J. Nelson, and D. B. Vigneron, "Flexible transceiver array for ultrahigh field human MR imaging," *Magn. Reson. Med.*, vol. 68, no. 4, pp. 1332–1338, Oct. 2012.

[22] M. J. Freire, M. A. Lopez, F. Meise, J. M. Algarin, P. M. Jakob, M. Bock, and R. Marques, "A Broadside-Split-Ring Resonator-Based Coil for MRI at 7 T," *IEEE Trans. Med. Imaging*, vol. 32, no. 6, pp. 1081–1084, Jun. 2013.

[23] Y. Li, B. Yu, Y. Pang, D. B. Vigneron, and X. Zhang, "Planar Quadrature RF Transceiver Design Using Common-Mode Differential-Mode (CMDM) Transmission Line Method for 7T MR Imaging," *PLOS ONE*, vol. 8, no. 11, p. e80428, Nov. 2013.

[24] C. H. Moon, J.-H. Kim, T. Zhao, and K. T. Bae, "Quantitative 23Na MRI of human knee cartilage using dual-tuned 1H/23Na transceiver array radiofrequency coil at 7 tesla," *J. Magn. Reson. Imaging*, vol. 38, no. 5, pp. 1063–1072, Nov. 2013.

[25] C. E. Akgun, L. DelaBarre, H. Yoo, S. M. Sohn, C. J. Snyder, G. Adriany, K. Ugurbil, A. Gopinath, and J. T. Vaughan, "Stepped Impedance Resonators for High-Field Magnetic Resonance Imaging," *IEEE Trans. Biomed. Eng.*, vol. 61, no. 2, pp. 327–333, Feb. 2014.

[26] X. Yan, R. Xue, and X. Zhang, "A monopole/loop dual-tuned RF coil for ultrahigh field MRI," *Quant. Imaging Med. Surg.*, vol. 4, no. 4, pp. 225–231, Aug. 2014.

[27] B. L. van de Bank, S. Orzada, F. Smits, M. W. Lagemaat, C. T. Rodgers, A. K. Bitz, and T. W. J. Scheenen, "Optimized 31P MRS in the human brain at 7 T with a dedicated RF coil setup," *NMR Biomed.*, vol. 28, no. 11, pp. 1570–1578, Nov. 2015.

[28] T. A. van der Velden, M. Italiaander, W. J. M. van der Kemp, A. J. E. Raaijmakers, A. M. T. Schmitz, P. R. Luijten, V. O. Boer, and D. W. J. Klomp, "Radiofrequency configuration to facilitate bilateral breast 31P MR spectroscopic imaging and high-resolution MRI at 7 Tesla," *Magn. Reson. Med.*, vol. 74, no. 6, pp. 1803–1810, Dec. 2015.

[29] J. Löring, W. J. M. van der Kemp, S. Almujayyaz, J. W. M. van Oorschot, P. R. Luijten, and D. W. J. Klomp, "Whole-body radiofrequency coil for 31P MRSI at 7 T," *NMR Biomed.*, vol. 29, no. 6, pp. 709–720, Jun. 2016.

[30] A. M. Nagel, R. Umathum, M. B. Rösler, M. E. Ladd, I. Litvak, P. L. Gor'kov, W. W. Brey, and V. D. Schepkin, "39K and 23Na relaxation times and MRI of rat head at 21.1 T," *NMR Biomed.*, vol. 29, no. 6, pp. 759–766, Jun. 2016.

[31] O. Rutledge, T. Kwak, P. Cao, and X. Zhang, "Design and test of a double-nuclear RF coil for 1H MRI and 13C MRSI at 7 T," *J. Magn. Reson.*, vol. 267, pp. 15–21, Jun. 2016.

[32] P. Cao, X. Zhang, I. Park, C. Najac, S. J. Nelson, S. Ronen, and P. E. Z. Larson, "1H-13C independently tuned radiofrequency surface coil applied for in vivo hyperpolarized MRI," *Magn. Reson. Med.*, p. n/a–n/a, Nov. 2015.

[33] F. Wetterling, M. Högler, U. Molkenthin, S. Junge, L. Gallagher, I. Mhairi Macrae, and A. J. Fagan, "The design of a double-tuned two-port surface resonator and its application to in vivo Hydrogen- and Sodium-MRI," *J. Magn. Reson.*, vol. 217, pp. 10–18, Apr. 2012.

[34] X. Yan, L. Shi, L. Wei, Y. Zhuo, X. J. Zhou, and R. Xue, "A hybrid sodium/proton double-resonant transceiver array for 9.4T MRI," in *Microwave Workshop Series on RF and Wireless Technologies for Biomedical and Healthcare Applications (IMWS-BIO), 2013 IEEE MTT-S International*, 2013, pp. 1–3.

[35] A. R. Rath, "Design and performance of a double-tuned bird-cage coil," *J. Magn. Reson. 1969*, vol. 86, no. 3, pp. 488–495, Feb. 1990.

[36] F. A. Sheikh, Y. C. Kim, I. C. Choi, and H. D. Kim, "Double-tuned RF receiver coil for detecting both 1H and 31P elements in 4.7 T MRI system," *Electron. Lett.*, vol. 51, no. 15, pp. 1150–1151, 2015.

[37] R. Brown, K. Lakshmanan, G. Madelin, and P. Parasoglou, "A nested phosphorus and proton coil array for brain magnetic resonance imaging and spectroscopy," *NeuroImage*, vol. 124, Part A, pp. 602–611, Jan. 2016.

[38] M. Augath, P. Heiler, S. Kirsch, and L. R. Schad, "In vivo 39K, 23Na and 1H MR imaging using a triple resonant RF coil setup," *J. Magn. Reson.*, vol. 200, no. 1, pp. 134–136, Sep. 2009.
[39] M. Rao, F. Robb, and J. M. Wild, "Dedicated receiver array coil for 1H lung imaging with same-breath acquisition of hyperpolarized 3He and 129Xe gas," *Magn. Reson. Med.*, vol. 74, no. 1, pp. 291–299, Jul. 2015.
[40] P. A. Bottomley, C. J. Hardy, P. B. Roemer, and O. M. Mueller, "Proton-decoupled, overhauser-enhanced, spatially localized carbon-13 spectroscopy in humans," *Magn. Reson. Med.*, vol. 12, no. 3, pp. 348–363, Dec. 1989.
[41] M. Alfonsetti, A. Sotgiu, and M. Alecci, "Design and testing of a 1.5 Tesla double-tuned (1H/31P) RF surface coil with intrinsic geometric isolation," *Measurement*, vol. 43, no. 9, pp. 1266–1276, Nov. 2010.
[42] W. m. Potter, L. Wang, K. k. McCully, and Q. Zhao, "Evaluation of a new 1H/31P dual-tuned birdcage coil for 31P spectroscopy," *Concepts Magn. Reson. Part B Magn. Reson. Eng.*, vol. 43, no. 3, pp. 90–99, Aug. 2013.
[43] A. Yahya, N. De Zanche, and P. S. Allen, "A dual-tuned transceive resonator for 13C{1H} MRS: two open coils in one," *NMR Biomed.*, vol. 26, no. 5, pp. 533–541, May 2013.
[44] M. A. Ohliger and D. K. Sodickson, "An introduction to coil array design for parallel MRI," *NMR Biomed.*, vol. 19, no. 3, pp. 300–315, May 2006.
[45] A. Kumar and P. A. Bottomley, "Optimized quadrature surface coil designs," *Magn. Reson. Mater. Phys. Biol. Med.*, vol. 21, no. 1–2, p. 41, Dec. 2007.
[46] G. Giovannetti, F. Frijia, S. Attanasio, L. Menichetti, V. Hartwig, N. Vanello, J. H. Ardenkjaer-Larsen, D. De Marchi, V. Positano, R. Schulte, L. Landini, M. Lombardi, and M. F. Santarelli, "Magnetic resonance butterfly coils: Design and application for hyperpolarized 13C studies," *Measurement*, vol. 46, no. 9, pp. 3282–3290, Nov. 2013.
[47] G. Giovannetti, F. Frijia, V. Hartwig, S. Attanasio, L. Menichetti, N. Vanello, V. Positano, J. h. Ardenkjaer-Larsen, V. Lionetti, G. d. Aquaro, D. D. Marchi, R. f. Schulte, F. Wiesinger, L. Landini, M. Lombardi, and M. f. Santarelli, "Design of a quadrature surface coil for hyperpolarized 13C MRS cardiac metabolism studies in pigs," *Concepts Magn. Reson. Part B Magn. Reson. Eng.*, vol. 43, no. 2, pp. 69–77, Apr. 2013.
[48] N. K. Bangerter, J. D. Kaggie, M. D. Taylor, and J. R. Hadley, "Sodium MRI radiofrequency coils for body imaging," *NMR Biomed.*, vol. 29, no. 2, pp. 107–118, Feb. 2016.
[49] G. Bogdanov and R. Ludwig, "Coupled microstrip line transverse electromagnetic resonator model for high-field magnetic resonance imaging," *Magn. Reson. Med.*, vol. 47, no. 3, pp. 579–593, Mar. 2002.
[50] B. Wu, X. Zhang, P. Qu, and G. X. Shen, "Design of an inductively decoupled microstrip array at 9.4 T," *J. Magn. Reson.*, vol. 182, no. 1, pp. 126–132, Sep. 2006.
[51] B. Wu, X. Zhang, P. Qu, and G. X. Shen, "Capacitively decoupled tunable loop microstrip (TLM) array at 7 T," *Magn. Reson. Imaging*, vol. 25, no. 3, pp. 418–424, Apr. 2007.

[52] K. Jasiński, A. Młynarczyk, P. Latta, V. Volotovskyy, W. P. Węglarz, and B. Tomanek, "A volume microstrip RF coil for MRI microscopy," *Magn. Reson. Imaging*, vol. 30, no. 1, pp. 70–77, Jan. 2012.

[53] J. Nilsson and S. Riedel, *Electric Circuits*. Pearson Higher Ed, 2014.

[54] J. C. Crane, M. P. Olson, S. J. Nelson, J. C. Crane, M. P. Olson, and S. J. Nelson, "SIVIC: Open-Source, Standards-Based Software for DICOM MR Spectroscopy Workflows, SIVIC: Open-Source, Standards-Based Software for DICOM MR Spectroscopy Workflows," *Int. J. Biomed. Imaging Int. J. Biomed. Imaging*, vol. 2013, 2013, p. e169526, Jul. 2013.

[55] C. Wuerslin, "Imagine - File Exchange - MATLAB Central," 2013. [Online]. Available: http://www.mathworks.com/matlabcentral/fileexchange/40440-imagine. [Accessed: 14-Aug-2016].

[56] G. W. Zack, W. E. Rogers, and S. A. Latt, "Automatic measurement of sister chromatid exchange frequency.," *J. Histochem. Cytochem.*, vol. 25, no. 7, pp. 741–753, Jul. 1977.

[57] A. M. Blamire, B. Rajagopalan, and G. K. Radda, "Measurement of myocardial pH by saturation transfer in man," *Magn. Reson. Med.*, vol. 41, no. 1, pp. 198–203, Jan. 1999.

[58] K. M. Brindle, B. Rajagopalan, D. S. Williams, J. A. Detre, E. Simplaceanu, C. Ho, and G. K. Radda, "31P NMR measurements of myocardial pH invivo," *Biochem. Biophys. Res. Commun.*, vol. 151, no. 1, pp. 70–77, Feb. 1988.

[59] A. Z. Lau, J. J. Miller, and D. J. Tyler, "Mapping of intracellular pH in the in vivo rodent heart using hyperpolarized [1-13C]pyruvate," *Magn. Reson. Med.*, p. n/a–n/a, Apr. 2016.

[60] B. H. Janssen, S. Lassche, M. T. Hopman, R. A. Wevers, B. G. M. van Engelen, and A. Heerschap, "Monitoring creatine and phosphocreatine by 13C MR spectroscopic imaging during and after 13C4 creatine loading: a feasibility study," *Amino Acids*, vol. 48, no. 8, pp. 1857–1866, Jul. 2016.

[61] Y. Li, Z. Xie, Y. Pang, D. Vigneron, and X. Zhang, "ICE decoupling technique for RF coil array designs," *Med. Phys.*, vol. 38, no. 7, pp. 4086–4093, Jul. 2011.

**Figure 1.** Proposed designed of the triple-nuclear coil. The loop coil, butterfly coil, and half-wavelength microstrip transmission line (MTL) are laid on top of each other.

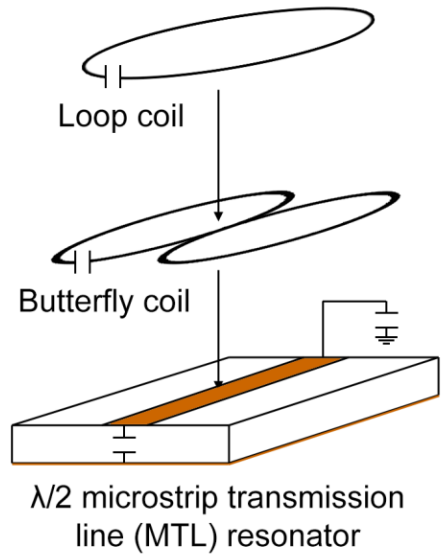

**Figure 2**. Top view (a) and bottom view (b) of the constructed coil. Each channel or resonant element has its own tuning and matching circuits.

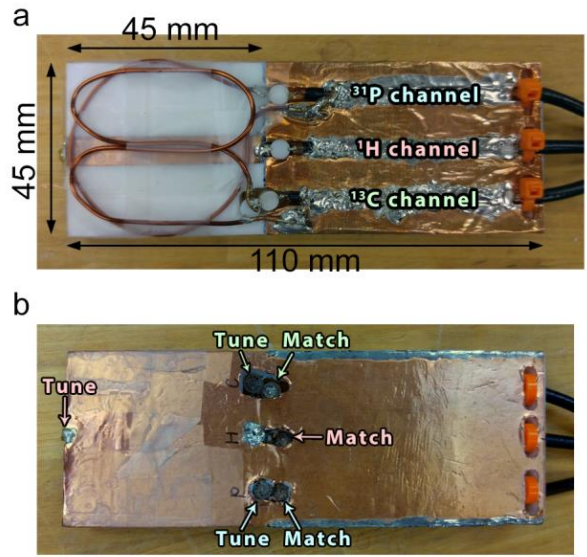

**Figure 3.** Coil test with the phosphate phantom. Proton images (a,c,e,g) and respective line profiles (b,d,f,h) along different planes and slices are shown. The dark bands in (e) and (g) are saturation bands inherent with the multi-slice and multi-plane acquisition. The raw (i) and apodized (j) spectra demonstrate feasibility of phosphorus acquisition.

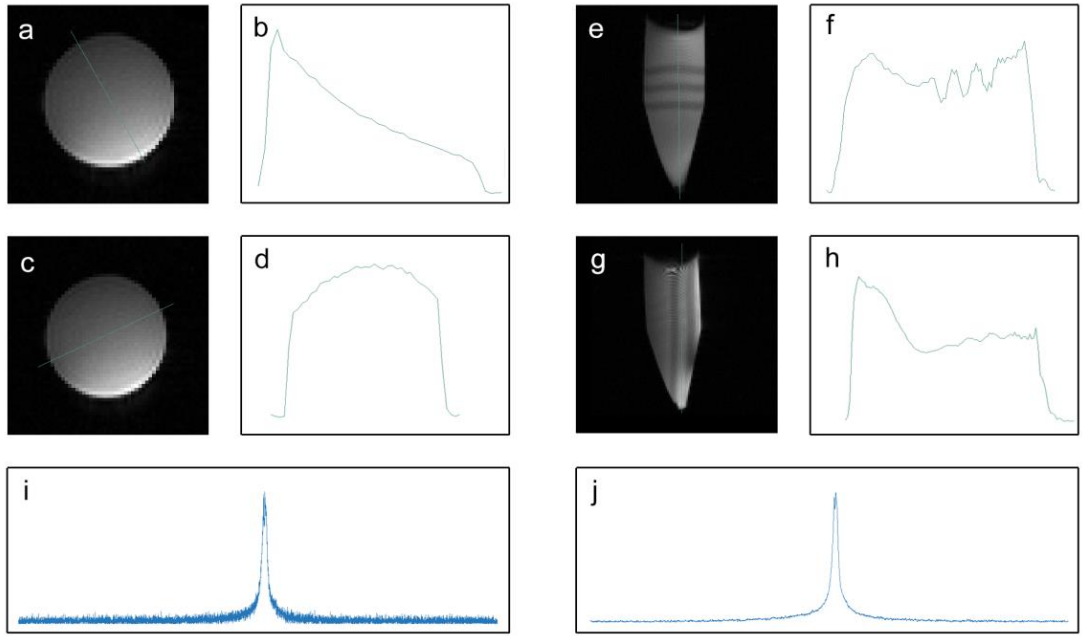

**Figure 4**. Coil test with $^{13}$C-Urea phantom. Proton images (a,c,e) and line profiles (b,d,f) along different planes and slices are shown. The MR spectra (h) are acquired from the slice shown in (g). The metabolite map (j) is overlaid on the proton image (i) using a sinc interpolator. A threshold of 2000 a.u. is used for the metabolite map.

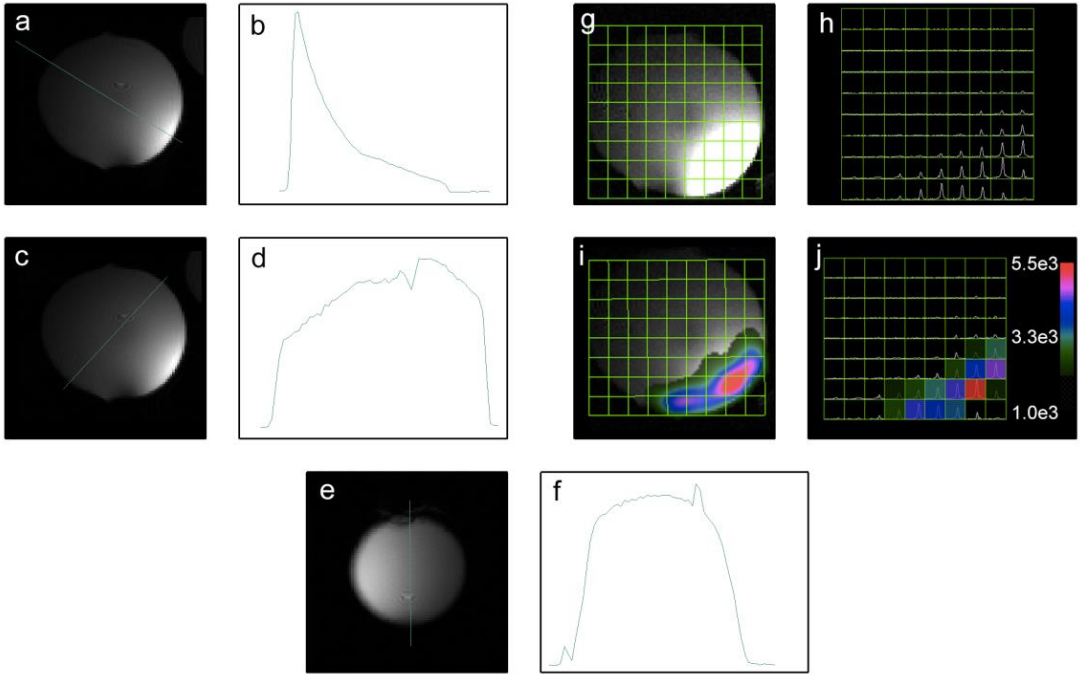

*Supplementary Material*

**Figure S1**. Plots of the $S_{11}$ (a) and $S_{21}$ (b) for the triple tuned coil. The coil achieves a worst-case $S_{11}$ of -27dB and worst-case $S_{21}$ of -16dB indicating good matching and decoupling performance.

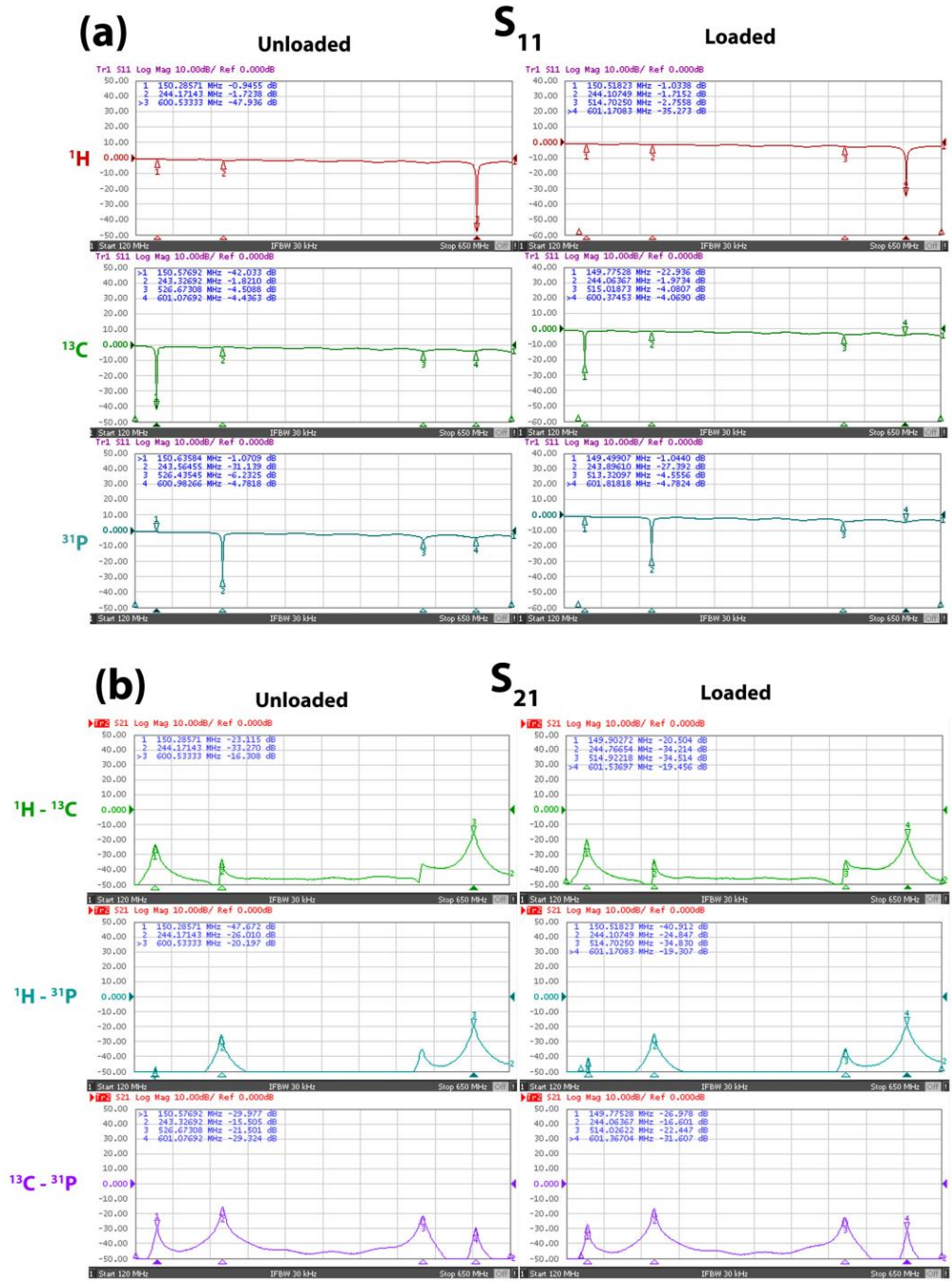